\documentclass[aps,prb,superscriptaddress,epsfigm,twocolumn,showpacs]{revtex4}
\usepackage{graphicx}
\usepackage{amsmath}

\begin{document}

\title{Effect of electron-electron interaction on the phonon-mediated
spin relaxation in quantum dots}

\author{Juan I. Climente}
\email{climente@unimore.it} \homepage{www.nanoscience.unimore.it}
\affiliation{CNR-INFM National Center on nanoStructures and bioSystems at Surfaces (S3), 
Via Campi 213/A, 41100 Modena, Italy}
\author{Andrea Bertoni}
\affiliation{CNR-INFM National Center on nanoStructures and bioSystems at Surfaces (S3), 
Via Campi 213/A, 41100 Modena, Italy}
\author{Guido Goldoni}
\affiliation{CNR-INFM National Center on nanoStructures and bioSystems at Surfaces (S3), 
Via Campi 213/A, 41100 Modena, Italy}
\affiliation{Dipartimento di Fisica, Universit\`a degli Studi di Modena e Reggio Emilia,
Via Campi 213/A, 41100 Modena, Italy}
\author{Massimo Rontani}
\affiliation{CNR-INFM National Center on nanoStructures and bioSystems at Surfaces (S3), 
Via Campi 213/A, 41100 Modena, Italy}
\author{Elisa Molinari}
\affiliation{CNR-INFM National Center on nanoStructures and bioSystems at Surfaces (S3), 
Via Campi 213/A, 41100 Modena, Italy}
\affiliation{Dipartimento di Fisica, Universit\`a degli Studi di Modena e Reggio Emilia,
Via Campi 213/A, 41100 Modena, Italy}
\date{\today}

\begin{abstract}

We estimate the spin relaxation rate due to spin-orbit coupling and
acoustic phonon scattering in weakly-confined quantum dots with up to five
interacting electrons.
The Full Configuration Interaction approach is used to account for the
inter-electron repulsion, and Rashba and Dresselhaus spin-orbit couplings 
are exactly diagonalized. We show that electron-electron interaction 
strongly affects spin-orbit admixture in the sample. Consequently, 
relaxation rates strongly depend on the number of carriers confined in the dot. 
We identify the mechanisms which may lead to improved spin stability in few 
electron ($>2$) quantum dots as compared to the usual one and two electron 
devices. Finally, we discuss recent experiments on triplet-singlet transitions 
in GaAs dots subject to external magnetic fields. Our simulations are in
good agreement with the experimental findings, and support the interpretation 
of the observed spin relaxation as being due to spin-orbit coupling assisted 
by acoustic phonon emission.

\end{abstract}


\maketitle


\section{Introduction}
\label{sec:intro}

There is currently interest in manipulating electron spins in quantum dots 
(QDs) for quantum information and quantum computing purposes.\cite{ZuticRMP,HeissSSC,LossPRA}
A major goal in this research line is to optimize the spin relaxation time ($T_1$),
which sets the upper limit of the spin coherence time ($T_2$): $T_2 \leq 2\, T_1$.\cite{GolovachPRL}
Therefore, designing two-level spin systems with long spin relaxation times
is an important step towards the realization of coherent quantum operations and 
read-out measuraments.
Up to date, spin relaxation has been investigated almost exclusively in 
single-electron\cite{GolovachPRL,HansonPRL1,KroutvarNAT,ElzermanNAT, Amasha_arxiv,KhaetskiiPRB,
ChengPRB,BulaevPRB,DestefaniPRB1,StanoPRB,WangPRB,ShermanPRB,WoodsPRB,
MerkulovPRB,ErlingssonPRB,SanJosePRL}
and two-electron\cite{FujisawaNAT,HansonPRL2,PettaSCI,JohnsonNAT,PettaPRB,
CoishPRB,FlorescuPEPRB,ChaneyPRB,ClimentePRBts,MeunierPRL,Golovach_arxiv} QDs.
Spin relaxation in QDs with a larger number of electrons has seldom been 
considered\cite{ClimentePRBts,SasakiPRL}, even though Coulomb blockade
makes it possible to control the exact number of carriers confined in a 
QD.\cite{CiorgaPRBDrexlerPRL}
Yet, recent theoretical works suggest that Coulomb interaction renders 
few-electron charge degrees of freedom more stable than single-electron 
ones\cite{BertoniPRLPRB}, which leads to the question of whether similar 
findings hold for spin degrees of freedom.
Moreover, in weakly-confined QDs, acoustic phonon emission assisted by spin-orbit 
(SO) interaction has been identified as the dominant spin relaxation mechanism 
when cotunneling and nuclei-mediated relaxation are 
reduced.\cite{KroutvarNAT,Amasha_arxiv,SasakiPRL} 
The combined effect of Coulomb interaction and SO coupling has been shown to
influence the energy spectrum of few-electron QDs profoundly,\cite{ChakrabortyPRB,
PietilainenPRB,DestefaniPRB2} but the consequences on the spin relaxation remain 
largely unexplored.\cite{aGolovach}

In Ref.~\onlinecite{ClimentePRBts} we investigated the effect of a magnetic field 
on the triplet-singlet (TS) spin relaxation in two and four-electron QDs with
SO coupling, so as to understand related experimental works.
Motivated by the very different response observed for different number of
confined particles, in this work we shall focus on the role
of electron-electron interaction in spin relaxation processes, extending
our analysis to different number of carriers, highlighting, in particular, 
the different physics involved in even and odd number of confined electrons.
Furthermore, we will explicitly compare the predictions of our theoretical 
model with very recent experiments on spin relaxation in two-electron 
GaAs QDs.\cite{MeunierPRL}

We study theoretically the energy structure and spin relaxation
of $N$ interacting electrons ($N=1-5$) in parabolic GaAs QDs with SO coupling, 
subject to axial magnetic fields. Both Rashba\cite{RashbaJPC} and 
Dresselhaus\cite{DresselhausPR} SO terms are considered, and the electron-electron 
repulsion is accounted for via the Full Configuration Interaction method.\cite{RontaniJPC,donrodrigo} 
By focusing on 
the two lowest spin states, two different classes of systems are distinguished. 
For $N$ odd (1,3,5) and weak magnetic fields, the ground state is a doublet and then the 
two-level system is defined by the Zeeman-split sublevels of the lowest orbital. 
For $N$ even (2,4), the two-level system is defined by a singlet and a triplet. 
We analyze these two classes of systems separately because, as we shall comment below, 
the physics involved in the spin transition differs. Thus, we compare 
the phonon-induced spin relaxation of $N=1,3,5$ electrons and that of $N=2,4$ separately.
As a general rule, the larger the number of confined carriers, the stronger the SO mixing, 
owing to the increasing density of electronic states. This would normally yield faster 
relaxation rates. However, we note that this is not necessarily 
the case, and few-electron states may display comparable or even slower relaxation 
than their single-electron and two-electron counterparts.
This is due to characteristic features of the few-particle energy spectra which 
tend to weaken the admixture between the initial and final spin states.
In $N$-odd systems, it is the presence of low-energy quadruplets for $N>1$ that 
reduces the admixture between the Zeeman sublevels of the (doublet) ground state,
hence inhibiting the spin flipping.
In $N$-even systems, electronic correlations partially quench phonon 
emission\cite{BertoniPRLPRB}, and the relaxation can be further suppressed
for $N>2$ if one selects initial and final spin states differing in more than 
one quantum of angular momentum, which inhibits direct triplet-singlet SO mixing
via linear Rashba and Dresselhaus SO terms.\cite{ClimentePRBts}
Noteworthy, all these effects are connected with Coulomb interaction between 
confined carriers.

The paper is organized as follows. In Section \ref{s:theo} we give details about 
the theoretical model we use. In Section \ref{s:odd} we study the energy structure 
and spin relaxation of a QD with an odd number of electrons ($N=1,3,5$).
In Section \ref{s:even} we do the same for QDs with an even number of electrons ($N=2,4$).
In Section \ref{s:exp} we compare our numerical simulations with experimental
data recently reported for $N=2$ GaAs QDs.
Finally, in Section \ref{s:conc} we present the conclusions of this work.

\section{Theory}
\label{s:theo}

We consider weakly-confined GaAs/AlGaAs QDs, which are the kind of 
samples usually fabricated by different groups to investigate spin
relaxation processes.\cite{ElzermanNAT,Amasha_arxiv,FujisawaNAT,PettaSCI}
In these structures, the dot and the surrounding barrier have similar
elastic properties, and the lateral confinement (which we approximate
as circular) is much weaker than the vertical one.
A number of useful approximations can be made for such QDs. First,
since the weak lateral confinement gives inter-level spacings within
the range of few meV, only acoustic phonons have significant
interaction with bound carriers, while optical phonons can be safely
neglected. Second, the elastically homogeneous materials are not
expected to induce phonon confinement, which allows us to consider
three-dimensional bulk phonons. Finally, the different energy scales of vertical
and lateral electronic confinement allow us to decouple vertical and
lateral motion in the building of single-electron spin-orbitals. Thus, we 
take a parabolic confinement profile in the in-plane $(x,y)$ direction,
with single-particle energy gaps $\hbar\omega_0$, which yields the
Fock-Darwin states.\cite{Pawel_book} In the vertical direction ($z$)
the confinement is provided by a rectangular quantum well of width
$L_z$ and height determined by the band-offset between the QD and 
barrier materials (the zero of energy is then the bottom of the
conduction band). The quantum well eigenstates are derived numerically. 
In cylindrical coordinates, the single-electron spin-orbitals 
can be written as:

\begin{equation}
\label{eq:0}
\psi_{\mu} (\rho,\theta,z;s_z) = 
\frac{1}{\sqrt{2\pi}}\,e^{i m \theta}\,R_{n,m}(\rho) \,\xi_0(z) \,\chi_{s_z},
\end{equation}

\noindent where $\xi_0$ is the lowest eigenstate of the quantum well, 
$\chi_{s_z}$ is the spinor eigenvector of the spin $z$-component
with eigenvalue $s_z$, and $R_{n,m}$ is the $n-$th Fock-Darwin orbital 
with azimuthal angular momentum $m$, 

\begin{equation}
\label{eq:1}
R_{n,m}(\rho)=\frac{1}{l_0}\,\sqrt{\frac{n!}{(n+|m|)!}}\,\left(\frac{\rho}{l_0}\right)^{|m|}\,
e^{-\frac{\rho^2}{2l_0^2}}\, {\cal L}_n^{|m|}\left(\frac{\rho^2}{l_0^2}\right).
\end{equation}

\noindent In the above expression ${\cal L}_n^{|m|}$ denotes a generalized 
Laguerre polynomial and $l_0=\sqrt{\hbar/m^* \omega_0}$ is the effective 
length scale, with $m^*$ standing for the electron effective mass. 
The energy of the single-particle Fock-Darwin states is given by
$E_{n,m}=(2\,n+1+\lvert m \rvert ) \hbar \Omega_c + \frac{m}{2}\,\hbar \omega_c$,
where $\omega_c=\frac{e B}{m^* c}$ is the cyclotron frequency and
$\Omega_c=\sqrt{\omega_0^2 + (\omega_c/2)^2}$ is the total (spatial plus
magnetic) confinement frequency.

With regard to Coulomb interaction, we need to go beyond mean field
approximations in order to properly include electronic correlations, which play an 
important role in determining the phonon-induced electron scattering rate.\cite{BraskenMP} 
Moreover, since we are interested in the relaxation time of excited states, we 
need to know both ground and excited states with comparable accuracy. Our method
of choice is the Full Configuration Interaction approach: the few-electron 
wave functions are written as linear combinations 
$|\Psi_a \rangle = \sum_i c_{a i} |\Phi_i\rangle$, where the Slater 
determinants $|\Phi_i \rangle=\Pi_{\mu_i} c_{\mu_i}^{\dagger} |0 \rangle$ 
are obtained by filling in the single-electron spin-orbitals $\mu$ 
with the $N$ electrons in \emph{all} possible ways consistent with symmetry 
requirements; here $c_{\mu}^{\dagger}$ creates an electron in the level
$\mu$. 
The fully interacting Hamiltonian is numerically diagonalized, 
exploiting orbital and spin symmetries.\cite{RontaniJPC,donrodrigo}
The few-electron states can then be labeled by the total azimuthal 
angular momentum $M=0,\pm 1,\pm 2\ldots$, total spin $S$ and its 
$z$-projection $S_z$.

The inclusion of SO terms is done following a similar scheme to that of
Ref. \onlinecite{LucignanoPRB}, although here we consider not only Rashba 
but also linear Dresselhaus terms.  
For a quantum well grown along the $[001]$ direction, these terms 
read:\cite{RashbaJPC,DresselhausPR}

\begin{eqnarray}
\label{eq:2}
{\cal H}^R &=& \frac{\alpha}{\hbar} (k_y s_x - k_x s_y),\\
\label{eq:2bis}
{\cal H}^D &=& \gamma_c\,\langle k_z^2 \rangle (k_y s_y - k_x s_x),
\end{eqnarray}

\noindent where $\alpha$ and $\gamma_c$ are coupling constants, 
while $s_j$ and $k_j$ are the $j$-th Cartesian projections of the electron 
spin and canonical momentum, respectively, along the main crystalographic axes
($\langle k_z^2 \rangle=(\pi/L_z)^2$ for the lowest eigenstate of 
the quantum well).
The momentum operator includes a magnetic field $B$ applied along the vertical 
direction $z$.
Other SO terms may also be present in the conduction band of a QD, 
such as the contribution arising from the system inversion asymmetry in the 
lateral dimension or the cubic Dresselhaus term.  However, in GaAs QDs with 
strong vertical confinement, ${\cal H}^R$ and ${\cal H}^D$ account
for most of the SO interaction.\cite{DestefaniPRB2}

We rewrite Eqs.(\ref{eq:2},\ref{eq:2bis}) in terms of ladder operators as:

\begin{eqnarray}
\label{eq:3}
{\cal H}^R = \frac{\alpha}{i \hbar} (k^+ s^- - k^- s^+),\\
\label{eq:3bis}
{\cal H}^D = \frac{\beta}{\hbar} (k^+ s^+ + k^- s^-),
\end{eqnarray}

\noindent where $k^{\pm}$ and $s^{\pm}$ change $m$ and $s_z$ by one quantum,
respectively, and $\beta=\gamma_c \, (\pi/L_z)^2$ is the Dresselhaus 
in-plane coupling constant. It is worth mentioning that when only Rashba (Dresselhaus)
coupling is present, the total angular momentum $j=m+s_z$ ($j=m-s_z$) is conserved.
However, in the general case, when both coupling terms are present and 
$\alpha \neq \beta$, all symmetries are broken.
Still, SO interaction in a large-gap semiconductor such as GaAs is rather weak, 
and the low-lying states can be safely labelled by their approximate quantum numbers 
$(M,S,S_z)$ except in the vicinity of the level anticrossings.\cite{BulaevPRB,FlorescuPEPRB,nohigh}
Since the few-electron $M$ and $S_z$ quantum numbers are given by the algebraic sum of the 
single-particle states $m$ and $s_z$ quantum numbers, it is clear from Eqs.~(\ref{eq:3},\ref{eq:3bis})
that Rashba interaction mixes $(M,S_z)$ states with $(M\pm1,S_z\mp1)$ ones, while 
Dresselhaus interaction mixes $(M,S_z)$ with $(M\pm1,S_z\pm1)$. 

The SO terms of Eqs.~(\ref{eq:3},\ref{eq:3bis}) can be spanned on a basis of correlated few-electron 
states.\cite{nopert} The SO matrix elements are then given by sums of single-particle 
contributions of the form:

\begin{multline}\label{eq:4}
\langle n'\, m'\, s_z'|\,{\cal H}^R+{\cal H}^D\,|n\, m\, s_z \rangle = \\
C_R^*\,{\cal O}^+_{n' m'\,n m}\,\delta_{m'\,m+1} \,\delta_{s_z'\,s_z-1}+
C_R  \,{\cal O}^-_{n'm'\,n m}\,\delta_{m'\,m-1} \, \delta_{s_z'\,s_z+1}+ \\
C_D^*\,{\cal O}^+_{n' m'\,n m}\,\delta_{m'\, m+1}\,\delta_{s_z'\,s_z+1} + 
C_D  \,{\cal O}^-_{n' m'\,n m}\,\delta_{m'\,m-1}\,\delta_{s_z'\,s_z-1}. 
\end{multline}

\noindent Here $C_R=\alpha$ and $C_D=-i \beta$ are constans for the Rashba and
Dresselhaus interactions respectively, and ${\cal O}^{\pm}$ are the form factors:

\small
\begin{multline}
\label{eq:5}
{\cal O}^-_{n' m'\,n m}= 
\frac{l_0}{2}  \int_0^{\infty}\,dt\, R_{n'm'}(t)
\left( 2\sqrt{t}\frac{\partial}{\partial t} + \frac{m}{\sqrt{t}} + \frac{B l_0^2 \sqrt{t}}{2} \right)
R_{nm}(t),
\end{multline}
\begin{multline}
\label{eq:6}
{\cal O}^+_{n' m'\,n m}= 
\frac{l_0}{2}  \int_0^{\infty}\,dt\, R_{n'm'}(t)
\left( 2 \left(\frac{\partial}{\partial t}\right)^\dagger \sqrt{t} + \frac{m'}{\sqrt{t}} + \frac{B l_0^2 \sqrt{t}}{2} \right)
R_{nm}(t),
\end{multline}
\normalsize

\noindent with $t=\rho^2/l_0^2$. 
The above forms factors have analytical expressions which depend 
on the set of quantum numbers $\{n'm', n m\}$. 
The resulting SO-coupled eigenvectors are then linear combinations of the 
correlated states, $|\Psi_A^{SO} \rangle = \sum_a c_{Aa} |\Psi_a\rangle$.

We assume zero temperature, which suffices to capture the main features of one-phonon 
processes.\cite{KhaetskiiPRB,WoodsPRB} Indeed, it is one-phonon processes that 
account for most of the low-temperature experimental observations in the SO 
coupling regime.\cite{HeissSSC,KroutvarNAT,Amasha_arxiv,ClimentePRBts,MeunierPRL,
SasakiPRL}
We evaluate the relaxation rate between the initial (occupied) and final (empty)
states of the SO-coupled few-electron state, B and A, using the Fermi Golden Rule:

\begin{multline}
\label{eq:7}
\tau^{-1}_{B \rightarrow A}=\frac{2\pi}{\hbar}\,\sum_{\nu \mathbf{q}}
\Bigl\lvert \sum_{ab} c_{Bb}^* c_{Aa} \sum_{ij} c_{bi}^* c_{aj} 
\langle \Phi_i|V_{\nu \mathbf{q}}|\Phi_j\rangle \Bigr\rvert^2\, 
\delta(E_B-E_A - \hbar \omega_q),
\end{multline}

\noindent where the electron states $| \Psi_K^{SO} \rangle$  ($K=\mbox{A,B}$) 
have been written explicitly as linear combinations of Slater determinants, 
$E_K$ stands for the $K$ electron state energy and $\hbar \omega_q$ represents 
the phonon energy.
$V_{\nu \mathbf{q}}$ is the interaction operator of an electron with an acoustic 
phonon of momentum $\mathbf{q}$ via the mechanism $\nu$, which can be either
deformation potential or piezoelectric field interaction.
Details about the electron-phonon interaction matrix elements can be found
elsewhere.\cite{BertoniPRLPRB}

In this work we study a GaAs/Al$_{0.3}$Ga$_{0.7}$As QDs, using
the following material parameters:\cite{Tin_book} electron effective
mass $m^*=0.067$, band-offset $V_c=243$ meV, crystal density
$d=5310$ kg/m$^3$, acoustic deformation potential constant $D=8.6$ eV,
effective dielectric constant $\epsilon=12.9$, and piezoelectric
constant $h_{14}=1.41\cdot 10^9$ V/m. The Land\'e factor is $g=-0.44$.\cite{HansonPRL1}
As for GaAs sound speed, we take $c_{\mbox{\tiny LA}}=4.72 \cdot 10^3$ m/s for 
longitudinal phonon modes and $c_{\mbox{\tiny TA}}=3.34 \cdot 10^3$ m/s 
for transversal modes.\cite{Landolt_book} 
Unless otherwise stated, a lateral confinement of $\hbar \omega_0=4$ meV 
and a quantum well width of $L_z=10$ nm are assumed for the QD under study,
and a Dressehlaus coupling parameter $\gamma_c=25.5$ eV$\cdot$\AA$^3$ is taken\cite{gammac}, 
so that $\beta \approx 25$ meV$\cdot$\AA. 
The value of the Rashba coupling constant can be modulated externally 
e.g.  with external electric fields. 
Here we will investigate systems both with and without Rashba interaction. 
When present, we shall mostly consider $\alpha=50$ meV$\cdot$\AA, to represent
the case where Rashba effects prevail over Dresselhaus ones.

Few-body correlated states $(M,S,S_z)$ are obtained using a 
basis set composed by the Slater determinants (SDs) which result from all possible 
combinations of 42 single-electron spin-orbitals (i.e., from the six lowest 
energy shells of the Fock-Darwin spectrum at $B=0$) filled with $N$ electrons. 
For $N=5$, this means that the basis rank may reach $\sim 2\cdot 10^5$.
The SO Hamiltonian is then diagonalized in a basis of up to 56 few-electron
states, which grants a spin relaxation convergence error below 2\%. 
Since SO terms break the spin and angular momentum symmetries, the
SO-coupled states $| \Psi_K^{SO} \rangle$ are described by a linear 
combination of SDs coming from different $(M,S,S_z)$ subspaces.
Thus, for $N=5$, the states are described by up to $\sim 8.5\cdot 10^5$ SDs.
To evaluate the electron-phonon interaction matrix elements, 
we note that only a small percentage of the huge number of possible pairs of 
SDs ($\sim 7\cdot10^{11}$ for $N=5$) may give 
non-zero matrix elements, owing to spin-orbital orthogonalities.
We scan all pairs of SDs and filter those which may 
give non-zero matrix elements writing the determinants in binary 
representation and using efficient bit-per-bit algorithms.\cite{RontaniJPC,donrodrigo}
The matrix elements of the remaining pairs ($\sim 2\cdot 10^6$ for $N=5$)
are evaluated using massive parallel computation.

\begin{figure}[h]
\includegraphics[width=5.5cm]{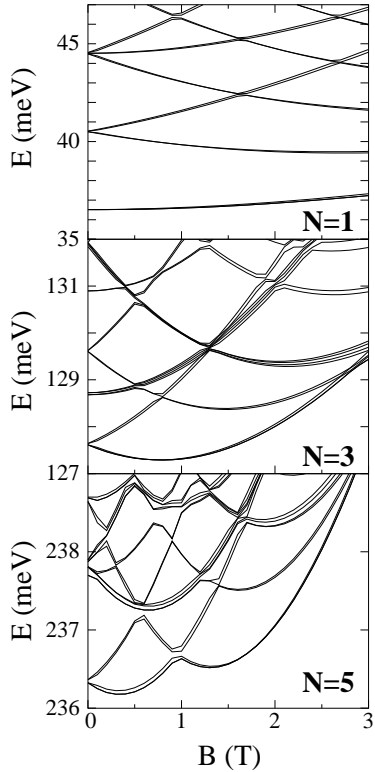}
\caption{Low-lying energy levels in a QD with $N=1,3,5$ interacting
electrons, as a function of an axial magnetic field.
The SO interaction coefficients are $\alpha=50$ meV$\cdot$ \AA~ 
and $\beta=25$ meV$\cdot$ \AA. The dot has $\hbar \omega_0=4$ meV
and $L_z=10$ nm.
Note the increasing size of the SO-induced anticrossing gaps
and zero-field splittings with increasing $N$.}\label{Fig1}
\end{figure}

\section{Spin relaxation in a QD with $N$ odd}
\label{s:odd}

\subsection{Energy structure}

When the number of electrons confined in the QD is odd and the 
magnetic field is weak enough, the ground and first excited states
are usually the Zeeman $s_z=1/2$ and $s_z=-1/2$ sublevels of a doublet
[Fig.~\ref{Fig1}]. 
Since the initial and final spin states belong to the same orbital,
$\Delta M=0$ and SO mixing (which requires $\Delta M=\pm 1$) 
is only possible with higher-lying states. In addition, the phonon
energy (corresponding to the electron transition energy) is typically 
small (in the $\mu$eV scale).
In this case, the relaxation rate is determined essentially by the 
phonon density, the strength and nature of the SO interaction, and 
the proximity of higher-lying states.\cite{KhaetskiiPRB,BulaevPRB}
In order to gain some insight on the influence of these
factors, in Fig.~\ref{Fig1} we compare the energy structure
of a QD with $N=1,3,5$ vs.~an axial magnetic field, in the presence of
Rashba and Dresselhaus interactions.\cite{fewe}
One can see that the increasing number of particles changes the energy
magneto-spectrum drastically. This is because the quantum numbers of the 
low-lying energy levels change, resulting in a different field dependence,
 and because Coulomb interaction leads to an increased density 
of electron states, as well as to a more complicated spectrum.

At first sight, the energy spectra of Fig.~\ref{Fig1} closely resemble
those in the absence of SO effects.
For instance, the $N=1$ spectrum is very similar to the pure Fock-Darwin
spectrum.\cite{Pawel_book}
Rashba and Dresselhaus interactions were expected to split the degenerate 
$|m|>0$ shells at $B=0$, shift the positions of the level crossings 
and turn them into anticrossings\cite{DestefaniPRB2,StanoPRB2,
VoskoboynikovPRB,KuanJAP}, but here such signatures are hardly visible 
because SO interaction is weak in GaAs. 
In fact, the magnitude of the SO-induced zero-field energy splittings and 
that of the anticrossing gaps is of very few $\mu$eV, and SO effects 
simply add fine features to the $N=1$ spectrum.\cite{StanoPRB2}

\begin{figure}[h]
\includegraphics[width=9.0cm]{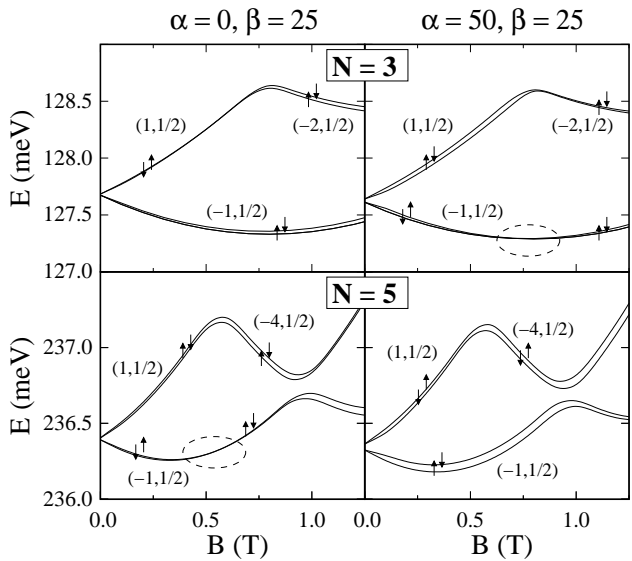}
\caption{The four lowest energy levels in a QD with $N=3,5$ interacting
electrons, as a function of an axial magnetic field, without (left column)
and with (right column) Rashba SO interaction. The approximate quantum 
numbers $(M,S)$ of the levels are shown, with arrows denoting the spin 
projection $S_z=1/2$ ($\uparrow$) and $S_z=-1/2$ ($\downarrow$). 
The dashed circles highlight the region of intrinsic spin mixing of 
the ground state.}\label{Fig2}
\end{figure}

A significantly different picture arises in the $N=3$ and $N=5$ cases.
Here, the increased density of electronic states enhances SO mixing as compared to
the single-electron case.\cite{enhance} As a result, the anticrossing gaps
can be as large as 30 $\mu$eV ($N=3$) and 60 $\mu$eV ($N=5$). 
Moreover, unlike in the $N=1$ case, where the ground state orbital 
has $m=0$, here it has $|M|=1$.  Therefore, the Zeeman sublevels involved 
in the fundamental spin transition are subject to SO-induced zero-field 
splittings. 
To illustrate this point, in Fig.~\ref{Fig2} we zoom in on the energy 
spectrum of the four lowest states of $N=3$ and $N=5$ under weak magnetic 
fields, without (left panels) and with (right panels) Rashba interaction.
Clearly, the four-fold degeneracy of $|M|=1$ spin-orbitals at $B=0$ 
has been lifted by SO interaction.\cite{DestefaniPRB2}
One can also see that the order of the two lowest sublevels 
at $B \sim 0$ changes when Rashba interaction is switched on.
Thus, for $N=3$ and $\alpha=0$, the two lowest sublevels
are $(M=-1,S_z=1/2)$ and $(M=-1,S_z=-1/2)$, but this order is reversed
when $\alpha=50$ meV$\cdot$\AA.
The opposite level order as a function of $\alpha$ is found for $N=5$.
This behavior constitutes a qualitative difference with respect to 
the $N=1$ case in two aspects.
First, the phonon energy (i.e., the energy of the fundamental
spin transition) is no longer given by the bare Zeeman splitting.
Instead, it has a more complicated dependence on the magnetic field, 
and it is greatly influenced by the particular values of 
$\alpha$ and $\beta$.
This is apparent in the $N=5$ panels, where the energy splitting 
between the two lowest states strongly differs depending on the
relative value of $\alpha$ and $\beta$.
Second, it is possible to find situations where the ground state 
at $B\sim 0$ has $S_z=-1/2$ and the first excited state has 
$S_z=1/2$ (e.g. $N=3$ when $\alpha > \beta$ or $N=5$ when 
$\alpha < \beta$). 
In these cases, the Zeeman splitting leads to a weak anticrossing
of the two sublevels (highlighted with dashed circles in
Fig.~\ref{Fig2}) which has no counterpart in single-electron 
systems. This kind of $B$-induced (i.e., not phonon-induced) 
ground state spin mixing, also referred to as 
``intrinsic spin mixing'', has been previously reported for 
singlet-triplet transitions in $N=2$ QDs.\cite{DestefaniPRB3} 
Here we show that they may also exist in few-electron QDs
with $N$ odd.

Figure \ref{Fig1} puts forward yet another qualitative difference 
between SO coupling in single- and few-electron QDs: while in 
the former low-energy anticrossings are due to Rashba 
interaction\cite{BulaevPRB,DestefaniPRB2,StanoPRB2},
in few-electron QDs, when $S=3/2$ states come into play, both
Rashba and Dresselhaus terms may induce anticrossings.
For example, the $(M=-1,S_z= 1/2)$ sublevel couples directly to 
both $(M=-2,S_z=-1/2)$ and $(M=-2,S_z=3/2)$ sublevels,
via the Dresselhaus and Rashba interaction, respectively.
Coupling to $S=3/2$ states is a characteristic feature
of $N>1$ systems, which has important effects on the spin
relaxation rate, as we will discuss below.

\subsection{Spin relaxation between Zeeman sublevels}

\begin{figure}[h]
\includegraphics[width=5.5cm]{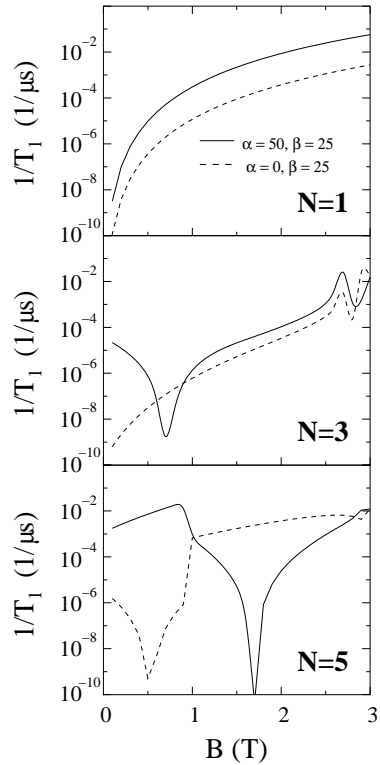}
\caption{Spin relaxation rate in a QD with $N=1,3,5$ interacting
electrons as a function of an axial magnetic field.
Solid (dashed) lines stand for the system with (without)
Rashba interaction. Note the strong influence of the
SO interaction in the shape of the relaxation curve
for $N>1$.}\label{Fig3}
\end{figure}

In Fig.~\ref{Fig3} we compare the magnetic field dependence of the
spin relaxation rate between the two lowest Zeeman sublevels of $N=1,3,5$. 
Dashed lines (solid lines) are used for systems without (with) Rashba 
interaction.\cite{b0}
While for $N=1$ the well-known exponential dependence with $B$ is 
found\cite{HeissSSC,KroutvarNAT,KhaetskiiPRB}, and the main effect of 
Rashba coupling is to shift the curve upwards (i.e., to accelerate the 
relaxation), for $N=3$ and $N=5$ the relaxation rate exhibits complicated 
trends which strongly depend on the values of the SO coupling parameters.

To understand this result, one has to bear in mind that in spin 
relaxation processes two well-distinguished and complementary
ingredients are involved, namely SO interaction and phonon emission. 
Phonon emission grants the conservation of energy in the electron relaxation,
but phonons have zero spin and therefore cannot couple states 
with different spin. It is the SO interaction that turns
pure spin states into mixed ones, thus enabling the 
phonon-induced transition. The overall efficiency of the scattering
event is then given by the combination of the two phenomena: the phonon
emission efficiency modulated by the extent of the SO mixing.
The shape of spin relaxation curves shown in Fig.~\ref{Fig3} 
can be directly related to the energy dispersion of the phonon,
which corresponds to the splitting between the two lowest levels of the 
electron spectrum.
Thus, for $N=1$, the phonon energy is simply proportional to $B$ through 
the Zeeman splitting, but for $N=3$ and $N=5$ it has a non-trivial 
dependence on $B$, as shown in Fig.~\ref{Fig2}. 
Actually, the relaxation minima in Fig.~\ref{Fig3} are connected with 
the magnetic field values where the two lowest levels anticross in 
Fig.~\ref{Fig2}. 
In these magnetic field windows, in spite of the fact
that SO coupling is strong, the phonon density is so small that the 
relaxation rate is greatly suppressed.\cite{ClimentePRBts}
Similarly, the relaxation rate fluctuations of $N=3$ at $B\sim 3$ T
are signatures of the anticrossings with high-angular momentum states.
For larger fields ($B>3$ T), the ground state approaches the maximum
density droplet configuration and high-spin states are 
possible.\cite{LucignanoPRB}
In this work, however, we restrict ourselves to the magnetic field regime 
where the ground state is a doublet. 

\begin{figure}[h]
\includegraphics[width=5.5cm]{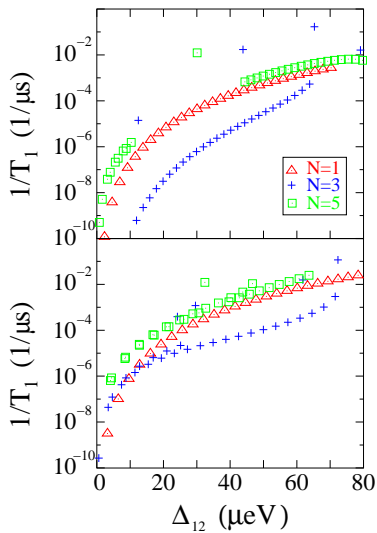}
\caption{(Color online). Spin relaxation rate in a QD with $N=1,3,5$ 
interacting electrons as a function of the energy splitting between the
two lowest spin states.
Top panel: $\alpha=0$, $\beta=25$ meV$\cdot$\AA.
Bottom panel: $\alpha=50$ meV$\cdot$\AA, 
$\beta=25$ meV$\cdot$\AA. The relaxation of $N=3$ is slower
than that of $N=1$ for a wide range of $\Delta_{12}$.
The irregular data distribution is due to the irregular 
relaxation rates vs.~magnetic field. For example, the strongly 
deviated points of $N=3$ come from the peaks at $B\sim3$ in 
Fig.~\ref{Fig3}.}\label{Fig4}
\end{figure}

For a more direct comparison between the relaxation rates of $N=1,3,5$,
in Fig.~\ref{Fig4} we replot the data of Fig.~\ref{Fig3} as a function of
the energy splitting between the two lowest states, $\Delta_{12}$, 
without (top panel) and with (bottom panel) Rashba interaction.
Since the phonon energy is identical for all points with the same $\Delta_{12}$, 
differences in the relaxation rate arise exclusively from the different 
strength of SO interaction.
$\Delta_{12}$ is also a relevant parameter from the experimental point of view,
since it is usually required that it be large enough for the states to be resolvable. 
In this sense, it is worth noting that, even if the inter-level splittings shown in 
Fig.~\ref{Fig4} are fairly small, a number of experiments have 
successfully addressed this regime.\cite{HansonPRL1,Amasha_arxiv,HansonPRL2}

A most striking feature observed in the figure is that, for most
values of $\Delta_{12}$, the $N=3$ relaxation rate is clearly slower than 
the $N=1$ one.
Likewise, $N=5$ shows a similar (or slightly faster) relaxation rate than $N=1$. 
These are interesting results, for they suggest that improved spin stability 
may be achieved using few-electron QDs instead of the single-electron ones 
typically employed up to date.\cite{Amasha_arxiv}
At first sight the results are surprising, because the higher density 
of states in the few-electron systems implies smaller inter-level spacings,
and hence stronger SO mixing, which should translate into enhanced relaxation.
It then follows that another physical mechanism must be acting upon the 
few-electron systems, which reduces the transition probability between the 
initial and final spin states, and may even make it smaller than for $N=1$.
Here we propose that such mechanism is the SO admixture with low-lying 
quadruplet ($S=3/2$) states, which become available for $N>1$.
By coupling to $S=3/2$ levels, the projection of the doublet
$S_z=1/2$ levels onto $S_z=-1/2$ ones is reduced, and this partly
inhibitis the transition between the lowest doublet sublevels.

Let us explain this by comparing the spin transition for $N=1$ and
$N=3$. For $N=1$, the spin configuration of the initial and final states, 
in the absence of SO coupling, is $\lvert S_z=-1/2 \rangle$ and 
$\lvert S_z=+1/2 \rangle$, respectively. The transition between these 
states is spin-forbidden. However, when SO coupling is switched on, the
two states become admixed with higher-lying $S=1/2$ states fulfilling 
the $\Delta S_z=\pm 1$ condition. The transition between the initial and
final states can then be represented schematically as:

\begin{equation}
c_a\,\lvert S_z=-1/2 \rangle + c_b  \lvert S_z=+1/2  \rangle \, \Rightarrow \,
c_r\,\lvert S_z=+1/2 \rangle + c_s  \lvert S_z=-1/2 \rangle,
\nonumber
\end{equation}

\noindent where $c_i$ are the admixture coefficients 
(in general $c_a\, \gg \, c_b$ and $c_r \, \gg \, c_s$).
Clearly now both spin configurations of the initial state have
a finite overlap with the final state, and so the transition is 
possible. Let us next consider the $N=3$ case. In the absence of
SO coupling, the initial and final states are again the
$S_z=-1/2$ and $S_z=+1/2$ doublets, respectively, and
the transition is spin-forbidden. When we switch on SO coupling,
we note that the $\Delta S_z=\pm 1$ condition allows for 
mixing not only with $S_z=\pm 1/2$ states (either doublets or 
quadruplets) but also with $S_z=\pm 3/2$ quadruplets, 
so that the transition can be represented as:

\begin{multline}
c_a  \lvert S_z=-1/2 \rangle +\, 
c_b  \lvert S_z=+1/2 \rangle + \,
c_c  \lvert S_z=-3/2 \rangle \,
\Rightarrow \, \\
c_r  \lvert S_z=+1/2 \rangle + \,
c_s  \lvert S_z=-1/2 \rangle + \,
c_t  \lvert S_z=+3/2 \rangle,
\nonumber
\end{multline}

\noindent where, in general, $c_a\, \gg \, c_b,\,c_c$,  and
$c_r \, \gg\, c_s,\,c_t$.
In this case, $\lvert S_z=-3/2 \rangle$ 
has no overlap with the final state configurations. Likewise, 
$\lvert S_z=+3/2 \rangle$ has no overlap with the initial 
state configurations.
Therefore, these quadruplet configurations are \emph{inactive} 
from the point of view of the transition, and the more
important they are (i.e., the stronger the SO coupling with quadruplet
states), the less likely the transition is. 

\begin{figure}[h]
\includegraphics[width=6.5cm]{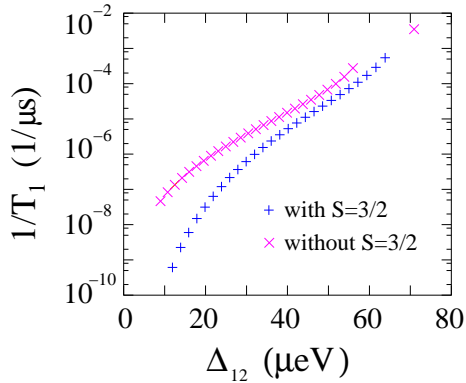}
\caption{(Color online). Spin relaxation rate in a QD with $N=3$ interacting
electrons as a function of the energy splitting between the
two lowest spin states. $\alpha=0$ and $\beta=25$ meV$\cdot$\AA.
Symbol $+$ ($\times$) stands for SO Hamiltonian diagonalized 
in a basis which includes (excludes) $S=3/2$ states.
Clearly, the inclusion of $S=3/2$ states slows down the 
relaxation.}\label{Fig5}
\end{figure}

To prove this argument quantitatively, in Fig.~\ref{Fig5} we illustrate 
the spin relaxation of $N=3$ calculated by diagonalization of the 
SO Hamiltonian including and excluding the low-lying $S=3/2$ states
from the basis set.
As expected, when the quadruplets are not considered, the transition 
is visibly faster.  
For $N=5$, low-lying $S=3/2$ levels are also available, but in this case
they barely compensate for the large density of electron states, so that the 
overall scattering rate turns out to be comparable to that of $N=1$.

\begin{figure}[h]
\includegraphics[width=5.5cm]{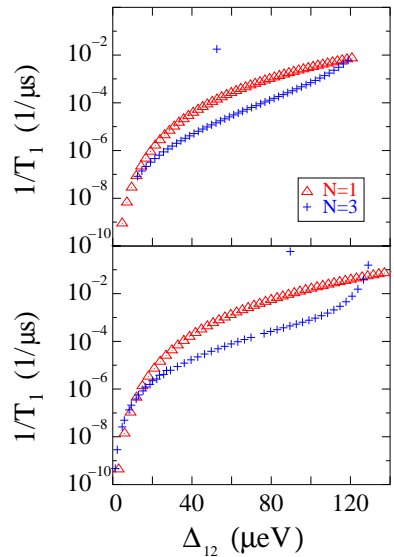}
\caption{(Color online). Spin relaxation rate in a QD with $N=1,3$ interacting
electrons as a function of the energy splitting between the
two lowest spin states. The QD has $\hbar \omega_0=6$ meV.
Top panel: $\alpha=0$, $\beta=25$ meV$\cdot$\AA.
Bottom panel: $\alpha=50$ meV$\cdot$\AA, 
$\beta=25$ meV$\cdot$\AA. As for the weaker-confined 
dot of Fig.~\ref{Fig4}, the relaxation of $N=3$ is slower than 
that of $N=1$ for a wide range of $\Delta_{12}$.}\label{Fig6}
\end{figure}

To test the robustness of the few-electron spin states
stability predicted above, we also compare the relaxation rate 
of $N=1$ and $N=3$ in a QD with different confinement, namely
$\hbar \omega_0=6$ meV, in Fig.~\ref{Fig6}.
Since the lateral confinement of the dot is now stronger,
$(M=-1,S=1/2)$ is the $N=3$ ground state up to large
values of the magnetic field ($B \sim 5$ T). This allows
us to investigate larger Zeeman splittings (i.e., larger
$\Delta_{12}$), which may be easier to resolve experimentally.
As seen in the figure, the relaxation rate of $N=3$ is again
slower than that of $N=1$ for a wide range of $\Delta_{12}$,
the behavior being very similar to that of Fig.~\ref{Fig4}, 
albeit extended towards larger inter-level spacings.
The crossing between $N=3$ and $N=1$ relaxation rates at
large $\Delta_{12}$ values, both in Fig.~\ref{Fig4} and 
Fig.~\ref{Fig6}, is due to the proximity of high-angular 
momentum levels coming down in energy for $N=3$ when the 
magnetic field (and hence the Zeeman splitting) is large.
Such levels bring about strong SO admixture and thus fast 
relaxation (see middle panel of Fig.~\ref{Fig3} at 
$B \sim 3$ T).

\section{Spin relaxation in a QD with $N$ even}
\label{s:even}

\subsection{Energy structure}

When the number of electrons confined in the QD is even and the 
magnetic field is not very strong, the ground and first excited states
are usually a singlet $(S=0)$ and a triplet $(S=1)$ with three Zeeman 
sublevels $(S_z=+1,0,-1)$.
Unlike in the previous section, here the initial and final states
of the spin transition may have different orbital quantum numbers,
and the inter-level splitting $\Delta_{12}$ may be significantly
larger (in the meV scale).
Under these conditions, the phonon emission efficiency no longer
exhibits a simple proportionality with the phonon density, 
but it further depends on the ratio between the phonon wavelength
and the QD dimensions.\cite{BockelmannPRB,ClimentePRB}
Moreover, SO interaction is sensitive to the quantum numbers 
of the initial and final electron states.\cite{FlorescuPEPRB,ClimentePRBts}
Therefore, in this class of spin transitions the details of the 
energy structure are also relevant to determine the relaxation rate.

\begin{figure}[h]
\includegraphics[width=5.5cm]{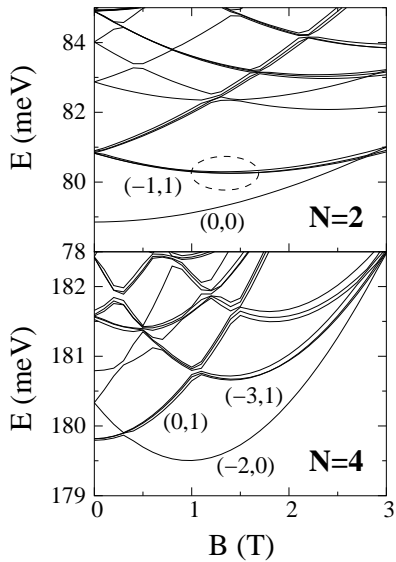}
\caption{Low-lying energy levels in a QD with $N=2,4$ interacting
electrons as a function of an axial magnetic field.
$\alpha=50$ meV$\cdot$ \AA~ and $\beta=25$ meV$\cdot$ \AA. 
The approximate quantum numbers
$(M,S)$ of the lowest states are shown. The dashed circle in $N=2$
highlights the anticrossing between $M=-1$ Zeeman sublevels.}\label{Fig7}
\end{figure}

In Fig.~\ref{Fig7} we plot the energy levels vs.~magnetic field
for a QD with $N=2,4$ in the presence of Rashba and Dresselhaus
interactions. The approximate quantum numbers $(M,S)$ of the lowest-lying 
states are written between parenthesis.
For $N=2$ and weak fields, the ground state is the $(M=0,S=0)$ singlet, 
and the first excited state is the $(M=-1,S=1)$  triplet.
As in the previous section, SO interaction introduces small zero-field 
splittings and anticrossings in the energy levels with $|M|>0$.\cite{DestefaniPRB2}
As a consequence, when $\alpha>\beta$, the zero-field ordering of the 
$(M=-1,S=1)$ Zeeman sublevels is such that they anticross in the presence
of an external magnetic field.  This anticrossing is highlighted in the figure
by a dashed circle.
On the other hand, as $B$ increases the singlet-triplet energy spacing is 
gradually reduced, and then the singlet experiences a series of weak 
anticrossings with all three Zeeman sublevels of the triplet.
These anticrossings are due to the fact that $(M=0,S=0,S_z=0)$ couples to 
the $(M=-1,S=1,S_z=-1)$ sublevel via Dresselhaus interaction, to the 
$(M=-1,S=1,S_z=+1)$ sublevel via Rashba interaction, and finally 
to the $(M=-1,S=1,S_z=0)$ sublevel indirectly through higher-lying 
states.\cite{FlorescuPEPRB,ClimentePRBts}

For $N=4$, the density of electronic states is larger than for $N=2$, 
which again reflects in a larger magnitude of the anticrossings 
gaps due to the enhanced SO interaction.
The ground state at $B=0$ is a triplet, $(M=0,S=1)$,
but soon after it anticrosses with a singlet, $(M=-2,S=0)$.
After this, and before the formation of Landau levels, 
two different branches of the first excited state can be 
distinguished: when $B<1$ T, the first excited state is 
$(M=0,S=1)$, and when $B>1$ T it is $(M=-3,S=1)$. 
It is worth pointing out that the complexity of the $N=4$ 
spectrum, as compared to the simple $N=2$ one, implies
a greater flexibility to select initial and final spin states
by means of external fields. 
As we shall discuss below, this degree of freedom has important
consequences on the relaxation rate.

\subsection{Triplet-singlet spin relaxation}

In a recent work, we have investigated the magnetic field dependence 
of the TS relaxation due to SO coupling and phonon emission in 
$N=2$ and $N=4$ QDs.\cite{ClimentePRBts}. 
Here we study this kind of transition from a different perspective,
namely we compare the spin relaxation of two- and four-electron systems 
in order to highlight the changes introduced by inter-electron repulsion.
Increasing the number of electrons confined in the QD has three 
important consequences on the TS transition. First, it increases 
the density of electronic states (and then the SO mixing), leading to faster
relaxation. Second, as mentioned in the previous section, it introduces 
a wider choice of orbital quantum numbers for the singlet and triplet states.
Third, it increases the strength of electronic correlations. 
Since now the initial and final spin states have different orbital
wave functions, the latter factor effectively reduces phonon scattering, 
in a similar fashion to charge relaxation processes\cite{BertoniPRLPRB}
(this effect has been recently pointed out in 
Ref.~\onlinecite{Golovach_arxiv} as well).
To find out the overall combined effect of these three factors, 
in this section we analyze quantitative simulations of correlated QDs.

\begin{figure}[h]
\includegraphics[width=5.5cm]{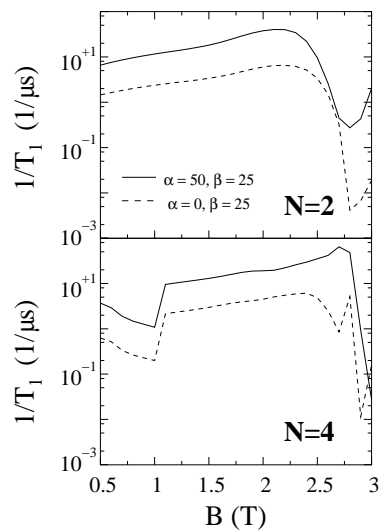}
\caption{Spin relaxation rate in a QD with $N=2,4$ interacting
electrons as a function of an axial magnetic field.
Solid (dashed) lines stand for the system with (without)
Rashba interaction. The relaxation of $N=4$ when $B<1$ T
is slower than that of $N=2$.}\label{Fig8}
\end{figure}

We focus on the magnetic field regions where the ground state 
is a singlet and the excited state is a triplet.
A complete description of the TS transition should then 
include spin relaxation between the Zeeman-split sublevels of
the triplet. However, for the weak fields we consider this 
relaxation is orders of magnitude slower than the TS one
(compare Figs.~\ref{Fig3} and \ref{Fig8}),\cite{antiphase}
the reason for this being the small Zeeman energy and the 
fact that the Zeeman sublevels are not directly coupled by 
Rashba and Dresselhaus terms, as mentioned in Section \ref{s:odd}.
Therefore, it is a good approximation to assume that all three 
triplet Zeeman sublevels are equally populated and they relax 
directly to the singlet.\cite{FlorescuPEPRB}

Figure \ref{Fig8} represents the TS relaxation rate in a QD with $N=2,4$, 
after averaging the relaxation from the three triplet sublevels.
Solid (dashed) lines stand for the case with (without) Rashba interaction.\cite{b0} 
The main effect of Rashba and Dresselhaus interactions is to accelerate
the spin transition by shifting the relaxation curve upwards.
This is in contrast to the $N$-odd case, where these terms may induce 
drastic changes in the shape of the relaxation rate curve (see Fig.~\ref{Fig3}).
Figure \ref{Fig8} also reveals a different behavior of the $N=2$ and 
$N=4$ TS relaxation rates.  The former increases gradually with $B$ and
then drops in the vicinity of the TS anticrossing, due to the small phonon
energies.\cite{ClimentePRBts,MeunierPRL,Golovach_arxiv}
Conversely, for $N=4$ an additional feature is found, namely an abrupt step
at $B \sim 1$. This is due to the change of angular momentum
of the excited triplet. For $B<1$ T the triplet has $M=0$, and 
for $B>1$ T it has $M=-3$. Since the ground state is a singlet with $M=-2$,
the $M=0$ triplet does not fulfill the $\Delta M=\pm 1$ condition for linear 
SO coupling. This inhibits direct spin mixing between initial and final states
and reduces the relaxation rate by about one order of magnitude.\cite{ClimentePRBts}

\begin{figure}[h]
\includegraphics[width=5.5cm]{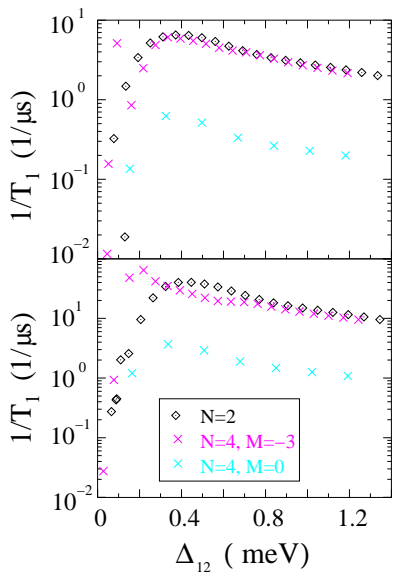}
\caption{(Color online). Spin relaxation rate in a QD with $N=2,4$ interacting
electrons as a function of the energy spacing between the singlet and the triplet.
Here $M$ stands for the angular momentum of the triplet.
Top panel: $\alpha=0$, $\beta=25$ meV$\cdot$\AA.
Bottom panel: $\alpha=50$ meV$\cdot$\AA, $\beta=25$ meV$\cdot$\AA. 
The relaxation of $N=4$ is comparable to that of $N=2$ when the triplet
has $M=-3$, and it is much smaller when $M=0$.}\label{Fig9}
\end{figure}

Noteworthy, the choice of states differing in more than one quantum of angular 
momentum is only possible for $N>2$ QDs. One may then wonder if it 
is more convenient to use these systems instead of the $N=2$ ones 
dominating the experimental literature up to date\cite{FujisawaNAT,
HansonPRL2,MeunierPRL}, i.e. if it compensates for the increased
density of electronic states. 
Interestingly, Fig.~\ref{Fig8} predicts slower relaxation for the $N=4$
QD with $M=0$ triplet than for $N=2$. To verify that this arises from
weakend SO coupling rather than from different phonon energy values,
in Fig.~\ref{Fig9} we replot the spin relaxation 
rate of $N=2,4$ as a function of the TS energy splitting.
In the figure, the upper and bottom panels represent the situations
without and with Rashba interaction, respectively.
While $N=4$ shows similar relaxation rate to $N=2$ when the triplet 
has $M=-3$, the relaxation is slower by about one order of magnitude 
when the triplet has $M=0$.
This result indicates that the weakening of SO mixing 
due to the violation of the $\Delta M=\pm 1$ condition
clearly exceeds the strengthening due to the higher density of states,
confirming that $N=4$ systems are more attractive than $N=2$ 
ones to obtain long triplet lifetimes.
We also point out that, in spite of the different density of states,
the relaxation rate of $N=2$ and $N=4,\,M=-3$ triplets is quite similar.
This can be ascribed to the phonon scattering reduction by electronic correlations,\cite{BertoniPRLPRB}
which may also explain the fact that experimentally resolved TS relaxation 
rates of $N=8$ QDs and $N=2$ QDs be quite similar.\cite{FujisawaNAT,SasakiPRL}

\section{Comparison with $N=2$ experiments}
\label{s:exp}

Whereas, to our knowledge, no experiments have measured transitions between Zeeman-split 
sublevels in $N>1$ systems yet, a number of works have dealt with TS 
relaxation in QDs with few interacting electrons. 
In Ref.~\onlinecite{ClimentePRBts} we showed that our model correctly
predicts the trends observed in experiments with $N=2$ and $N=8$ QDs
subject to axial magnetic fields.\cite{FujisawaNAT,HansonPRL2,SasakiPRL}
In this section, we extend the comparison to new experiments available 
for $N=2$ TS relaxation in QDs,\cite{MeunierPRL} which for the first time 
provide continuous measurements of the average triplet lifetime against axial
magnetic fields, from $B=0$ to the vicinity of the TS anticrossing. 
By using a simple model, the authors of the experimental work showed that 
the measuraments are in clear agreement with the behavior expected from 
SO coupling plus acoustic phonon scattering.
However, in such model: (i) the TS energy splitting was a taken directly from
the experimental data, (ii) the SO coupling effect was accounted for by parametrizing
the admixture of the lowest singlet and triplet states only, and (iii) the 
$B$-dependence of the SO-induced admixture was neglected.
Approximation (ii) may overlook the correlation-induced reduction
of phonon scattering,\cite{BertoniPRLPRB,Golovach_arxiv} that we have shown
above to be significant, and which may have an important contribution from 
higher excited states in weakly-confined QDs.
In turn, approximation (iii) may overlook the important influence of SO coupling 
in the $B$-dependence of the triplet lifetime, as we had anticipated in 
Ref.~\onlinecite{ClimentePRBts}.
Here we compare with the experimental findings using our model, which 
includes these effects properly. We assume a QD with an effective
well width $L_z=30$ nm, as expected by Ref.~\onlinecite{MeunierPRL} authors,
and a lateral confinement parabola of $\hbar \omega_0=2$ meV
which, as we shall see next, fits well the position of the TS anticrossing.
Yet, the comparison is limited by the lack of detailed information about the 
Rashba and Dresselhaus interaction constants, and because we deal with 
circular QDs instead of elliptical ones (the latter effect introduces simple 
deviations from the circular case\cite{FlorescuPEPRB}).
In addition, in the experiment a tilted magnetic field of magnitude $B^*$,
forming an angle of $68^\circ$ with the vertical direction was used. Here
we consider the vertical component of the field ($B=0.37\,B^*$), 
which is the main responsible for the changes in the energy structure, 
and the effect of the in-plane component enters via the Zeeman splitting only.

\begin{figure}[h]
\includegraphics[width=5.3cm]{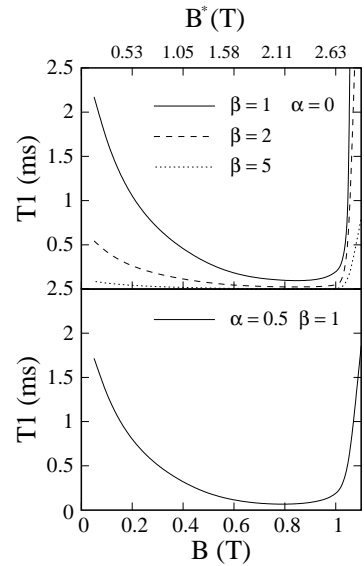}
\caption{Average triplet lifetime in a QD with $N=2$ electrons as a function of 
an axial magnetic field. Only the field region before the TS anticrossing is shown.
$\alpha$ and $\beta$ are in meV$\cdot$\AA~units. $B$ is the applied axial magnetic
field, and $B^*$ is the equivalent tilted magnetic field, for comparison with
Ref.~\onlinecite{MeunierPRL} experiment.}\label{Fig10}
\end{figure}

Figure \ref{Fig10} illustrates the average triplet lifetime for $N=2$.
The bottom axis shows the vertical magnetic field $B$ value, while
the top axis shows the value to be compared with the experiment $B^*$.\cite{b0}
As can be seen, the triplet lifetime first decreases with the field and 
then it abruptly increases in the vicinity of the TS anticrossing, due to 
the small phonon density.\cite{ClimentePRBts}
This behavior is in clear agreement with the experiment (cf.~Fig.~3 of 
Ref.~\onlinecite{MeunierPRL}).
The position of the anticrossing ($B^* \sim 2.9$ T) is also close to the
experimental value ($B^* \sim 2.8$ T), which confirms that that 
$\hbar \omega_0=2$ meV is similar to the mean confinement frequency of the 
experimental sample.
A departure from the experimental trend appears at weak fields ($B<0.5$ T),
where we observe a continuous increase of $T_1$ with decreasing $B$, while the 
experiment reports a plateau.
This is most likely due to the ellipticity of the experimental sample, which renders 
the electron states (and consequently the relaxation rate) insensitive to 
the field in the $B^*=0-0.5$ T region (see Fig.~1a in \onlinecite{MeunierPRL}).
In any case, Fig.~\ref{Fig10} clearly confirms the role of phonon-induced relaxation 
in the experiments, using a realistic model for the description of correlated 
electron states, SO admixture and phonon scattering.

A comment is worth here on the magnitude of the SO coupling terms. 
In Fig.~\ref{Fig10}, we obtain good agreement with the experimental relaxation 
times by using small values of the SO coupling parameters.
In particular, a close fit is obtained using $\beta=1,\,\alpha=0.5$ meV$\cdot$\AA, 
which yields a spin-orbit length $\lambda_{SO} = 48$ $\mu$m. This value, 
which coincides with the experimental guess ($\lambda_{SO} \approx 50$ $\mu$m), 
indicates that SO coupling is several times weaker than that reported for 
other GaAs QDs.\cite{Amasha_arxiv} 
Typical GaAs parameters are often larger. For instance, measuraments 
of the Rashba and Dresselhaus constants by analysis of the weak antilocalization in 
clean GaAs/AlGaAs two-dimensional gases revealed $\alpha=4-5$
meV$\cdot$\AA, and $\gamma_c=28$ eV$\cdot$\AA$^3$ (i.e, $\beta=3$ meV$\cdot$\AA$\,$
for our quantum well of $L_z=30$ nm).\cite{MillerPRL}
To be sure, the small SO coupling parameters in the experiment have
a major influence on the lifetime scale. Compare e.g. the $\beta=1$ and 
$\beta=5$ meV$\cdot$\AA$\,$ curves in Fig.~\ref{Fig10}. 
Actually, we note that accurate comparison with the timescale reported for
other GaAs samples\cite{SasakiPRL} is also possible within our model, 
but assuming stronger SO coupling constants.\cite{ClimentePRBts}
In Ref.~\onlinecite{MeunierPRL}, it was suspected that the weak SO coupling
inferred from the experimental data could be the result of the exclusion of 
higher orbitals and the magnetic field dependence of SO admixture in their 
model (higher states reduce the \emph{effective} SO coupling constants 
by decreasing the phonon-induced scattering\cite{Golovach_arxiv,BertoniPRLPRB}).
Here we have considered both these effects and still small SO coupling
constants are needed to reproduce the experiment. Therefore,
understanding the origin of their small value remains as an open 
question. One possibility could be that the particular direction of the
tilted magnetic field used in the experiment corresponded to a reduced 
degree of SO admixture.\cite{Golovach_arxiv}

\section{Conclusions}
\label{s:conc}

We have investigated theoretically the energy structure and spin relaxation 
rate of weakly-confined QDs with $N=1-5$ interacting electrons, subject to axial
magnetic fields, in the presence of linear Rashba and Dresselhaus SO interactions.
It has been shown that the number of electrons confined in the dot introduces
changes in the energy spectrum which significantly influence the intensity of
the SO admixture, and hence the spin relaxation. In general, the larger the
number of confined carriers, the higher the density of electronic states. This decreases
the energy splitting between consecutive levels and then enhances SO admixture,
which should lead to faster spin relaxation.
However, we find that this is not necessarily the case, and slower relaxation
rate may be found for few-electron QDs as compared to the usual single and 
two-electron QDs used up to date.  
The physical mechanisms responsible for this have been identified.
For $N$-odd systems, when the spin transition takes place between Zeeman-split
sublevels, it is the presence of low-energy $S=3/2$ states for $N>1$ that 
reduces the projection of the doublet $S_z=1/2$ sublevels into $S_z=-1/2$ ones, 
thus partly inhibiting the spin transition.
For $N$-even systems, when the spin transition takes place between triplet and
singlet levels, there are two underlying mechanisms. 
On the one hand, electronic correlations tend to reduce phonon emission efficiency.
On the other hand, for $N>2$ a magnetic field can be used to select a pair 
of singlet-triplet states which do not fulfill the $\Delta M=\pm 1$ condition of
direct SO admixture, which significantly weakens the SO mixing.

Last, we have compared our estimates with recent experimental data for TS 
relaxation in $N=2$ QDs.\cite{MeunierPRL} Our results support the interpretation of 
the experiment in terms of SO admixture plus acoustic phonon scattering, even though 
quantitative agreement with the experiment requires assuming much weaker SO coupling 
than that reported for similar GaAs structures.\\

\begin{acknowledgments}
We acknowledge support from the Italian Ministry for University and
Scientific Research under FIRB RBIN04EY74, Cineca Calcolo parallelo 2006,
and Marie Curie IEF project NANO-CORR MEIF-CT-2006-023797.
\end{acknowledgments}


\begin{thebibliography}{61}
\bibitem{aGolovach}
During the finalization of this paper we have learned about a parallel
work investigating the influence of Coulomb interaction in two-electron
TS relaxation.\cite{Golovach_arxiv} Many of the findings in such paper
are in agreement with our numerical results.
\bibitem{nohigh}
The $(M,S,S_z)$ quantum numbers of few-electron states are a good 
approximation for the lowest-lying states only. For higher-lying
states, the energy spectrum becomes denser and the
SO interaction becomes very strong even for GaAs, which leads to
important departures from the SO-free picture. This does not occur
in single-electron parabolic QDs because the energy levels are
equally spaced.
\bibitem{nopert}
The convenience of using exact diagonalization procedures, instead 
of perturbational approaches, to account for the SO coupling in GaAs 
QDs has been claimed in Ref.~\onlinecite{ChengPRB}.
\bibitem{fewe}
The energy magneto-spectrum of GaAs parabolic QDs with SO 
interaction and up to four interacting electrons was 
also investigated in Ref.~\onlinecite{PietilainenPRB}, but  
considering Rashba interaction only.
\bibitem{enhance} 
Coulomb-enhanced SO interaction was previously predicted for 
higher-dimensional structures.\cite{ChenPRB} 
Here we report it for QDs.
\bibitem{b0}
For simplicity of the discussion, in Figs.~\ref{Fig3}, \ref{Fig8} 
and \ref{Fig10}, the near vicinity of $B=0$ T is not shown. In
that range one finds damped phonon-induced relaxation rates due
to degeneracies arising from the time-reversal symmetry and the
circular symmetry of the confinement we have assumed. We do not
expect these features to be observable in experiments, because QDs 
are not perfectly circular and because hyperfine interaction is 
expected to be the dominant spin relaxation mechanism for very 
weak fields (see Refs.~\onlinecite{ErlingssonPRB},\onlinecite{JohnsonNAT}).
\bibitem{antiphase}
Greatly suppressed TS spin relaxation, comparable to that of inter-Zeeman 
sublevels at very weak $B$, may be achieved by means of geometrically 
or field-induced acoustic phonon emission minima.\cite{ChaneyPRB,ClimentePRBts}
\end{thebibliography}
\end{document}